\documentclass[amsmath,amssymb,amsfonts,aps,prl]{revtex4}

\usepackage{graphicx}
\begin{document}
\title{Evolution of surname distribution under gender-equality measurements}
\author{L. F. Lafuerza, R. Toral}
\affiliation{IFISC, Instituto de F{\'\i}sica Interdisciplinar y Sistemas Complejos, CSIC-UIB,  Campus UIB, E-07122 Palma de Mallorca, Spain}
\date{\today}
\begin{abstract}
We consider a model for the evolution of the surnames distribution under a gender-equality measurement presently discussed in the Spanish parliament (the children take the surname of the father or the mother according to alphabetical order). We quantify how this would bias the alphabetical distribution of surnames, and analyze its effect on the present distribution of the surnames in Spain.
\end{abstract}

\maketitle
\section{Introduction}
In Spain, as in many other countries, children usually inherit the surname of the father. As a consequence, the surname of the mother is lost in the children's generation\cite{spain}. Nowadays, in Spain, parents can agree upon whether it is the mother's or the father's surname that is given to their children, but if parents do not reach an agreement, it will be the father's surname the one inherited by the children. Due to gender-equality issues, a new law is under study which would imply that, if parents do not reach an agreement, or if no wish is expressed, the surname inherited by the children will be selected according to the alphabetical order of the parent's two surnames.

People have immediately realized that this implies a bias on the surnames favoring those beginning by the first letters in the alphabet (A,B,\dots) and could mean the disappearance of surnames beginning by the last letters (\dots,Y,Z). In this short note, we quantify the effect of this bias on the present distribution of surnames in Spain. 
\section{Model}
As a first order model that captures the essence of the process of surname inheritance we propose the following:\newline
$(i)$ Initially, a population of $2N$ individuals ($N$ male and $N$ female) is considered. Each individual has a surname chosen according to some prescribed distribution.\newline
$(ii)$ Males and females reproduce in random pairs in such a way that, on average, the total population remains constant. \newline 
$(iii)$ With probability $a$ it is assumed that parents reach an agreement, so that the surnames of the children are chosen at random between those of the parents (it is not important for the results in which proportion they prefer the father's or the mother's surname). With probability $1-a$, parents do not reach or do not express an agreement, and the children adopt the surname by the alphabetical order rule. \newline
We measure time $t$ in average reproductions per person, or generations.  In a generation, parents are replaced by their children in the population.

This is a  minimal model and does not consider many realistic issues: new surnames brought in by immigration, geographical distribution of surnames, etc. but those are expected to be second order effects with little impact in the overall trend.

Let us define $p(n,t)$ as the proportion of individuals (both males and females) with surname in the alphabetical position $n=1,\dots,M$, being $M$ the total number of surnames. It evolves according to:
\begin{equation}
 \frac{\partial p(n,t)}{\partial t}=(1-a)p(n,t)\left[\sum_{k=n+1}^Mp(k,t) - \sum_{k=1}^{n-1}p(k,t)\right]=(1-a)p(n,t)\left[1-P(n,t)-P(n-1,t)\right]\label{eq1},
\end{equation}
where $P(n,t)=\sum_{k=1}^np(k,t)$ is the cumulative distribution. It follows that:
\begin{equation}
\frac{\partial P(n,t)}{\partial t}=(1-a)P(n,t)\left[1-P(n,t)\right],
\label{eq2}
\end{equation}
whose solution is:
\begin{equation}
 P(n,t)=\frac{P(n,0)e^{(1-a)t}}{1+P(n,0)(e^{(1-a)t}-1)}.\label{P(t)}
\end{equation}
The distribution of surnames at time $t$ is then $p(1,t)=P(1,t)$ and $p(n,t)=P(n,t)-P(n-1,t)$ if $n>1$. Approximating the difference by a derivative $p(n,t)\simeq\frac{\partial P(n,t)}{\partial t}$, we obtain:
\begin{equation}
 p(n,t)=\frac{p(n,0)e^{(1-a)t}}{\left[1+P(n,0)(e^{(1-a)t}-1)\right]^2}\label{p(t)}.
\end{equation}
Eq. (\ref{P(t)}) shows that the distribution of surnames approaches a Kronecker-delta at $n=1$ ($P(n,t)=1, \forall n$) exponentially fast with a characteristic time $1/(1-a)$. Assuming, for instance, that couples reach and agreement about the children's name and express it in $50\%$ of the cases ($a=1/2$), we find from Eq. (\ref{p(t)}) that the frequency of a surname around the end of the alphabetical table would be decreased by a factor $10$ in around $4.6$ generations($\sim115$ years). 

\section{Evolution of current distribution}
We have applied the above results to the actual distribution of Spanish surnames. Besides the analytical result of Eq,(\ref{p(t)}), we have performed a numerical simulation of the model in which $N=10^7$ couples have probabilities $(0.05,0.2,0.5,0.2,0.05)$ of having $(0,1,2,3,4)$ children (average value is $2$). The probability of parents reaching an agreement is set at $a=0.5$. Independently on whether an agreement has been reached or not, the rule applied to the first-born child is used for all children. We have used as the initial condition $p(n,0)$ the distribution of the $M=100$ most common surnames in Spain, as published by the INE~\cite{ine}, after ordering them by alphabetical order. The evolution after $n=4$ and $n=10$ generations is plotted in the figure. The agreement between the simulation and the analytical result is excellent. 

\section{Conclusions}
In our minimal model for surname transmission, we prove that the adoption of the alphabetical rule leads to an exponential decrease  for the surnames in the last positions in the alphabetical order, with a characteristic decay time of $1/(1-a)$ generations, begin $a$ the fraction of parents that reach an agreement, This quantifies the decrease in the frequency of those surnames.
\newline
\newline
\newline
\includegraphics[width=16.5cm, height=5.0cm, clip=true]{figure.eps}
\newline
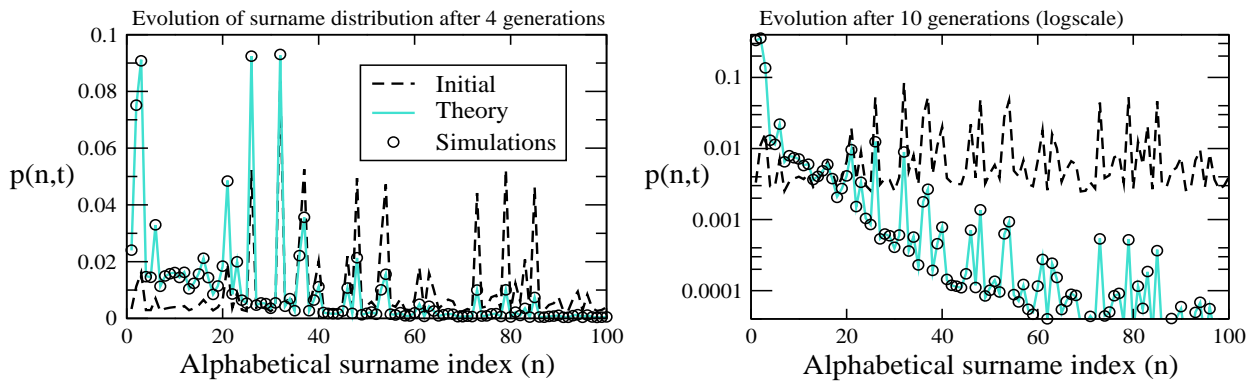
{Fig.1.-\small{Evolution of the distribution of surnames after $n=4$ (left) and $n=10$ generations, taking as initial condition $p(n,0)$ the actual distribution of the $M=100$ most common surnames in Spain.  For $n=10$ we have used a logarithmic scale for a better viewing of the data. The dots are the result of the numerical simulation of a more detailed model that includes the basic premises used in the derivation of the analytical expression. \label{fig}}
\newline
\newline
{\textbf{Acknowledgments:}}
We acknowledge financial support by the MC (Spain) and FEDER (EU) through project FIS2007-60327. L.F.L. is supported by the JAEPredoc program of CSIC.


\begin{thebibliography}{99}
\bibitem{spain} In Spain, though, the mother's surname is kept as a second surname. It is consequently totally lost in the grand-children's generation.
\end{thebibliography}
\end{document}